\documentclass[a4paper,onecolumn,unpublished]{quantumarticle}
\pdfoutput=1

\usepackage[T1]{fontenc}
\usepackage[utf8]{inputenc} 
\usepackage{amsmath}

\usepackage{hyperref} 
\usepackage[english]{babel}

\usepackage{tikz} 

\usepackage[marginal]{footmisc} 

\usepackage{ulem} 

\newcommand{\orcidlink}[1]{\href{https://orcid.org/#1}{\includegraphics[width=8pt]{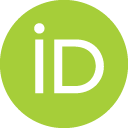}}}

\begin{document}

\title{Predicting RNA Secondary Structure on Universal Quantum Computer}

\author[1]{Ji Jiang$^{\#,}$}
\author[1]{Qipeng Yan$^{\#,}$}
\author[1]{Ye Li}
\author[1]{Min Lu}
\author[1]{Ziwei Cui}
\author[1]{Menghan Dou}
\author[2]{Qingchun Wang\orcidlink{0000-0003-0966-9782}$^{*,}$}
\author[2,3,4]{Yu-Chun Wu\orcidlink{0000-0002-8997-3030}}
\author[1,2,3,4]{Guo-Ping Guo\orcidlink{0000-0002-2179-9507}$^{*,}$}
\footnote{\noindent ${^\#}$These authors contributed equally to this work. \\
*Correspondence authors: qingchun720@ustc.edu.cn and gpguo@ustc.edu.cn}

\affiliation[1]{Origin Quantum Computing Company Limited, Hefei, Anfei 230026, China}
\affiliation[2]{Institute of Artificial Intelligence, Hefei Comprehensive National Science Center, Hefei, Anfei 230088, China}
\affiliation[3]{CAS Key Laboratory of Quantum Information, School of Physics, University of Science and Technology of China, Hefei 230026, China}
\affiliation[4]{CAS Center for Excellence in Quantum Information and Quantum Physics, University of Science and Technology of China, Hefei 230026, China}

\maketitle

\begin{abstract}
It is the first step for understanding how RNA structure folds from base sequences that to know how its secondary structure is formed. Traditional energy-based algorithms are short of precision, particularly for non-nested sequences, while learning-based algorithms face challenges in obtaining high-quality training data. Recently, quantum annealer has rapidly predicted the folding of the secondary structure, highlighting that quantum computing is a promising solution to this problem. However, gate model algorithms for universal quantum computing are not available. In this paper, gate-based quantum algorithms will be presented, which are highly flexible and can be applied to various physical devices. Mapped all possible secondary structure to the state of a quadratic Hamiltonian, the whole folding process is described as a quadratic unconstrained binary optimization model. Then the model can be solved through quantum approximation optimization algorithm. We demonstrate the performance with both numerical simulation and experimental realization. Throughout our benchmark dataset, simulation results suggest that our quantum approach is comparable in accuracy to classical methods. For non-nested sequences, our quantum approach outperforms classical energy-based methods. Experimental results also indicate our method is robust in current noisy devices. It is the first instance of universal quantum algorithms being employed to tackle RNA folding problems, and our work provides a valuable model for utilizing universal quantum computers in solving RNA folding problems.
\end{abstract}

\section{INTRODUCTION}
Ribonucleic Acid (RNA) is a biological macromolecule with a complex three-dimensional spiral folding structure, which is exactly in this especial structure that many important cellular processes are carried out, such as catalytic reactions, regulation of gene expression, regulation of innate immunity, and sensing of tiny molecules~\cite{higgs2000rna, cooper2009rna, cech2014noncoding, wang2012splicedisease}. Therefore, understanding of RNA structure is crucial for biological engineering especially in designing synthetic RNA, discovering RNA-targeted drugs, genome editing and vaccine development~\cite{mathews2006revolutions, warner2018hajdin, churkin2018design, sharp2009centrality}.

Cellular RNA is usually a single strand composed of four kinds of bases (\textit{i.e.} A, C, G, and U) connected by phosphoric diester bond. These bases can also pair with each other through hydrogen bonds, which is referred to as base pairing, such as canonical base pairing (A-U, C-G)~\cite{sloma2017base}, non-Watson-Crick pairing (G-U)~\cite{fallmann2017recent}, and non-canonical base pairing~\cite{westhof2000rna}. So, single-stranded RNA can fold into more complex structure. The single strand, that is the sequence information of bases, is the primary structure of RNA. The secondary structure is a hairpin-shaped composite structure formed by convolutional folding of the primary structure. The tertiary structure is a spatial structure established by further bending the spiral secondary structure. Finally, the mixture of nucleic acid and protein due to the interaction between RNA and protein is considered as the quaternary structure. It is noteworthy that, in many cases, the secondary structure is often in a more dominant position than the sequence itself. For example, the secondary structure in many homologous types of RNA bear significant resemblance among divergent sequences~\cite{fox19755s, mathews2010folding}. Furthermore, a correct secondary structure is also the cornerstone of a stable tertiary and quaternary structure~\cite{mortimer2014insights}. Therefore, secondary structure is the foundation of the whole complex structure and acquiring its detailed information is vital for RNA research.

Structural biologists have developed a variety of experimental methods to obtain the secondary structure of RNA, such as dimethyl sulfate mutational profiling with sequencing (DMS-MaPseq)~\cite{cordero2012quantitative}, Selective 2’-hydroxyl acylation analyzed by primer extension (SHAPE)~\cite{wilkinson2006selective, mortimer2007fast}, \textit{X}-ray crystallography~\cite{furtig2003nmr}, nuclear magnetic resonance (NMR)~\cite{cheong2004rapid}, and cryogenic electron microscopy~\cite{fica2017cryo}. However, these experimental methods share common drawbacks of high cost, low yield, and low throughput, making it difficult for large scale application.

On the other hand, the improved capabilities of digital electronic computers allow computational biologists to predict secondary structure based on primary structure under an abstract model, the RNA folding problem. Nevertheless, predicting RNA secondary structure remains a huge challenge today due to vast solution space in mathematics. Most of the classical algorithms~\cite{mathews2006revolutions, fallmann2017recent, reuter2010rnastructure, zhao2021review} find the structure with a minimized thermodynamic free energy through dynamic programming, which has been adapted by the famous Vienna RNAfold~\cite{lorenz2011viennarna}, MFold~\cite{zuker2003mfold} and RNAstructure~\cite{mathews2006prediction}. But, they have no guaranteed performance when the RNA contains non-nested pattern called pseudoknot. Using dynamic programming to predict a proper structure for such types of RNA with  is NP-complete~\cite{lyngso2000rna}. In other words, it is generally believed that no classical algorithm can find the optimal solution within a moderate time. Recently, due to the great success of deep learning in protein structure prediction~\cite{du2021trrosetta, jumper2021highly, humphreys2021computed, wang2016protein}, similar methods have been explored for RNA folding problem~\cite{zhang2019new, wang2019dmfold, chen2020rna,singh2019rna, fu2022ufold} but they are less effective for RNA. This is not only because the availability of RNA template structure database is low but also because the sequence convolution information of RNA is less helpful for the algorithm to obtain the structural contact characteristics~\cite{pucci2020evaluating}.

With the continuous development of quantum computing, a number of quantum algorithms emerge and show remarkable achievements in classically intractable problems. In particular, some recent works have been done concerning prediction of RNA secondary structure using quantum annealer (QA)~\cite{fox2022rna, zaborniak2022qubo}. However, study related to gate model is scarcely mentioned in literature. Gate-based devices can perform universal quantum computation using quantum gates, providing improved algorithm design flexibility. For example, the Shor’s algorithm and Grover’s algorithm are both beyond the reach of QA~\cite{9781107002173}. 

In this paper, we will develop gate model quantum simulation algorithms to address this issue, which is a universal quantum algorithm that runs on gate-based quantum devices and is more flexible in application to various physical hardware. Specifically, we first establish a quadratic unconstrained binary optimization (QUBO) model to describe the folding process of RNA where the optimal structure is encoded as the ground state of the quadratic Hamiltonian. Then the model is solved with quantum approximation optimization algorithm (QAOA) introduced by Farhi \textit{et al.}~\cite{farhi2014quantum}. QAOA is a heuristic algorithm for combinatorial optimization problems, and has a provable performance enhancement over classical algorithms in some hard problems~\cite{crooks2018performance}. We will present an X-mixers QAOA to solve the RNA folding problem by means of noiseless digital simulation. Our results suggest that this quantum algorithm can achieve a high success rate in sampling optimal structures. In instances with pseudoknots, this quantum algorithm outperforms classical energy-based methods. More importantly, we verify this algorithm on real quantum computers through cloud computing platform. Despite of noise and limited connectivity of physical qubits, X-mixers still provide more than one half chance to measure the ground state in most cases. Furthermore, an alternative module for QAOA, \textit{i.e.} parity-partitioned $XY$-mixers, is suggested. This algorithm shows promise as a quantum approach for large-scale RNA sequences by directly implementing the hidden structure of the problem.

\section{MATERIALS AND METHODS}

\subsection{QUBO formulation of RNA folding problem}

In this section, we introduce the QUBO formulation of the RNA folding problem and present a phenomenological objective function to evaluate RNA structures, which provides a friendly approach for gate-based algorithms.

\begin{figure*}
    \centering
    \includegraphics[width=\textwidth]{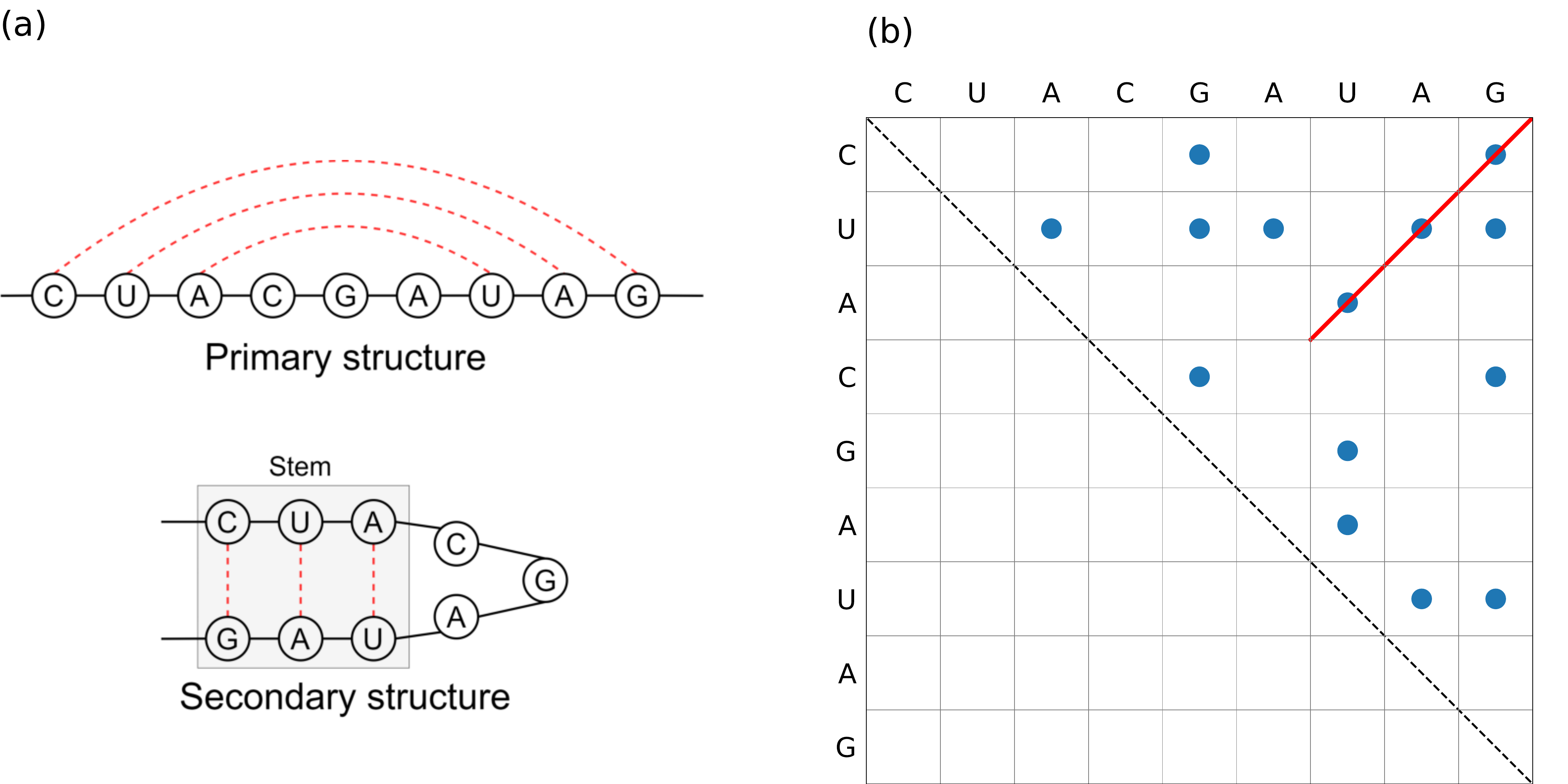}
    \caption{\label{fig:rna_stems} Visualization of (a) the  structure of the sample sequence and (b) its pairing matrix. The sequence is presented as a string of beads with the red dashed lines indicating hydrogen-bond interactions of the stem. We only plot the upper triangular part of the pairing matrix given the fact that it must be symmetric.}
\end{figure*}

Before looking into the QUBO model of the RNA folding problem, it is necessary to establish a mathematical representation of the RNA secondary structure. In this paper, we only consider canonical base-pairs (A-U and C-G) and non-Watson-Crick base-pairs (G-U). Take the sequence CUACGAUAG in Fig. \ref{fig:rna_stems}(a) as an example. If a string of bases to describe its primary structure are used, the secondary structure is given by a set of consecutive base-pairs, which is often referred to as stems. The number of base-pairs in a stem is defined as its length. Visualizing the pairing matrix can certainly provide another intuitive perspective. By denoting possible base-pairs (A-U, C-G and G-U) by 1 and illegal pairing by 0, a pairing matrix presents all base-pairs of the RNA. As shown in Fig. \ref{fig:rna_stems}(b), we use colored dots to highlight possible base-pairs of the example sequence ('1's in the matrix). Since the stems must be successive base-pairs by definition, it is easy to recognize them as strings of consecutive dots in the direction perpendicular to the main diagonal (dashed line). To obtain a stable structure with strong interaction, the minimum stem length is adapted as 3. For our example sequence, there is only one such stem which has been labelled by the red line in the pairing matrix. Note that all stems can be found within quadratic time by visiting each element of the matrix~\cite{kai2019efficient}.

Suppose $N$ stems have been obtained from the matrix: $S=\{ s_i | i=1, 2, ..., N \}$. Any secondary structure must be one of its subset. The RNA folding problem is to find a subset $A \subseteq S$ corresponding to the optimal structure. If a string of bit-wise variables $\vec{x}=x_{1}x_{2}...x_{N-1}x_{N}$ is defined such that $x_i=1$ if $s_i \in A$ otherwise $x_i=0$, then our goal is to search for the string representing the optimal structure from $2^N$ possible combinations. The next step is to design an objective function $C(\vec{x})=C(x_1, x_2, ..., x_N)$ which determines whether a set of stems is the structure we are looking for. The bit-wise variables $\vec{x}$ and the objective function $C(\vec{x})$ explicitly define the QUBO model. Inspired by several classical models~\cite{lewis2021qfold, kai2019efficient}, we design a linear increasing model which grants efficient implementation for gate-based quantum computers. The goal aims to maximize the number of base-pairs with as few stems as possible, meanwhile making sure no stem overlaps. Since only the consecutive base pairs would contribute to the reduction of free energy in most energy-based models~\cite{kai2019efficient}, this approach also captures certain physical essence of RNA structure. It leads to the following quadratic function:
\begin{equation}
    C(\vec{x}) = \sum_{i} 2 k_i x_i - \sum_{i} \frac{N_{seq}}{2 k_i + \epsilon} x_i + \sum_{i} \sum_{j<i} K_{ij} x_i x_j,
    \label{eq:obj_func}
\end{equation}
where $k_i$ is the length of $s_i$, $N_{seq}$ is the number of bases of the sequence, and $\epsilon$ is a tunable hyperparameter which we will discuss later. $K_{ij}$ is a penalty function for overlapping stems and pseudoknots written as:
\begin{equation}
    K_{ij} =
    \begin{cases}
      -(k_i + k_j) & \text{if $s_i$ and $s_j$ are overlapping}\\
      c_p (k_i + k_j) & \text{if $s_i$ and $s_j$ form a pseudoknot}\\
      0 & \text{otherwise}
    \end{cases}
    .
    \label{eq:kij}
\end{equation}

The leading terms in Eq. (\ref{eq:obj_func}) sums the total number of base-pairs contributed by all selected stems while the third term cancels half of the contribution if $s_i$ and $s_j$ are overlapping. Combined they aim to maximize the number of base-pairs with a combination of non-overlapping stems. If two stems form a pseudoknot, the value of parameter $c_p$ in $K_{ij}$ is tunable where $1 < c_p < 0$ if we favor the non-nest structure, $-1 < c_p < 0$ if we discourage it, or $c_p = 0$ if we simply want degenerate states combing structures both with and without pseudoknots. Here we let $c_p=0$ so that both structures may be sampled by the quantum algorithm. The remaining second term is designed to be an effective penalty to the number of stems. In a sequence with $N_{seq}$ bases, there can exist $\frac{N_{seq}}{2 k}$ stems at most assuming all stems have the same length. However, a stable structure should spare some bases unpaired so that the molecular may not rupture itself. We introduce the hyperparameter $\epsilon$ to describe such flexibility, which acts as the averaged free bases of the sequence (the total number of unpaired bases divided by the maximum number of possible stems). Therefore, the sequence can hold no more than the number of $\frac{N_{seq}}{2 k_i + \epsilon}$ length-$k$ stems at any circumstances. With these terms subscribed, each stem inherits a certain amount of penalty and the shorter is it the larger is the penalty. Specifically, the objective function will prefer choosing one long stem than multiple short stems with an equal number of bases. For instance, it will favor one length-6 stem over two length-3 stems. We empirically set $\epsilon=6$ in this manuscript.

\subsection{Quantum computing and QAOA framework}

\begin{figure}
    \centering
    \includegraphics[width=\columnwidth]{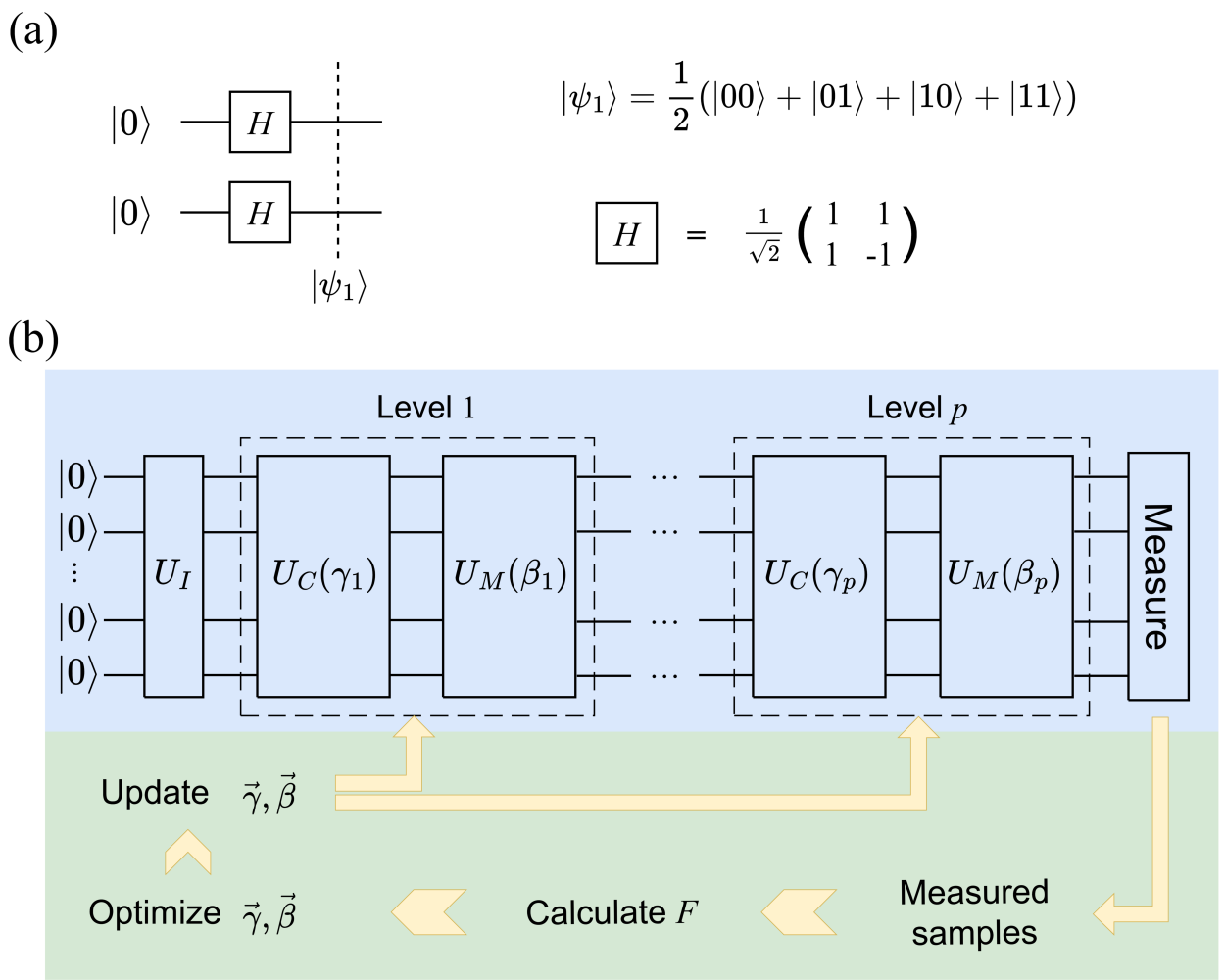}
    \caption{\label{fig:qaoa_frame} (a) The quantum circuit to prepare a superposition of all possible states in a two-qubit system. On the right hand side the final state and the matrix representation of Hadamard gate are written explicitly. (b) The general framework of QAOA. It is a hybrid algorithm running on both quantum computers (light blue parts) and classical computers (light green parts). The modules $U_I$, $U_C(\gamma)$ and $U_M(\beta)$ are modules made up of multiple quantum gates.}
\end{figure}

Quantum computing is a cutting-edge paradigm which has unparalleled advantages over classical computations in certain problems. The quantum analogue uses \textit{qubit}, which is intrinsically a two-level quantum state, as the fundamental unit of information. In this paper we employ Dirac notation and use $| \cdot \rangle$ to denote a quantum state. The vital difference between a qubit and a classical bit is that the qubit is able to stay in a superpositioned state due to its quantum nature. In a less rigorous sense, a qubit can be 0 and 1 \textit{simultaneously} during calculations. Theoretically, quantum computer allows us to perform calculations on exponentially vast amounts of states ($2^N$) by manipulating only $N$ qubits. This counter-intuitive behavior grants us access to amazing quantum resources such as quantum parallel and quantum entanglement. In particular, some recent works have been done concerning prediction of RNA secondary structure using quantum annealer~\cite{fox2022rna, zaborniak2022qubo}. However, study related to gate model is scarcely mentioned in literature. Gate-based devices can perform universal quantum computation using quantum gates, providing improved flexibility for algorithm design 
which is beyond the reach of quantum annealer. 
Quantum gates, which are similar to the concept of classical logic gates, are used to manipulate the information stored in qubits. The calculations are carried out by a series of quantum gates which is conveniently presented by quantum circuits. We give a canonical example in Fig. \ref{fig:qaoa_frame}(a) where a superposition of all possible states with two Hadamard gates is prepared for a two-qubit system. If the resulting states $| \psi _1 \rangle$ are measured, an equal chance to be obtained one of the classical results: 00, 01, 10 and 11.

QAOA is a hybrid quantum-classical algorithm designed to run a gate-based devices and to find approximate solutions to combinatorial optimization problems, attracting extensive attention in recent years~\cite{PhysRevA.97.022304, pagano2020quantum, PhysRevResearch.2.023074, PhysRevApplied.14.034009, PhysRevX.10.021067, harrigan2021quantum, PhysRevResearch.4.033029, zhang2022quantum, niroula2022constrained}. Especially the \textit{quantum alternating operator ansatz} extended by Halfield \textit{et al.}~\cite{hadfield2019quantum} allows more flexible choice of certain modules, providing convenient ways to implement hard constraints such as those in Max-$\kappa$-Colorable-Subgraph problem~\cite{PhysRevA.101.012320} and Maximum \textit{k}-Vertex Cover problem~\cite{cook2020quantum}. Fig. \ref{fig:qaoa_frame}(b) illustrates the general framework of QAOA. It has two parameterized modules referred to as cost layer $U_C(\gamma)$ and mixer layer $U_M(\beta)$, where $\gamma$ and $\beta$ are real numbers. QAOA applies the two modules alternatively for $p$ times to a easy-to-prepare state $|\psi \rangle = U_I | 0 \rangle ^{\otimes n}$ which can be prepared with polynomial quantum gates. The final state is given by:
\begin{equation}
   |\phi \rangle = U_M(\beta _p) U_C(\gamma _p) ... U_M(\beta _1) U_C(\gamma _1) |\psi \rangle,
\end{equation}
where the integer $p$ is defined as the level of QAOA. The exact form of the cost layer is problem-dependent. In fact, it simulates a Hamiltonian $H_C$ obtained from the objective function by simply replacing the binary variables $x_i$ in Eq. \ref{eq:obj_func} with $(I-Z_i)/2$ where $Z_i$ is the Pauli-$Z$ gate acting on the $i$-th qubit. With $H_C$ given, the cost layer has the following form:
\begin{equation}
    U_C(\gamma)=e^{-i \gamma H_C},
\end{equation}
where
\begin{equation}
\begin{aligned}
    H_C = & -\sum_{i}  2 k_i \frac{I-Z_i}{2} + \sum_{i} \frac{N_{seq}}{2 k_i + \epsilon} \frac{I-Z_i}{2}  \\ & - \sum_{i} \sum_{j<i} K_{ij} \frac{I-Z_i}{2} \frac{I-Z_j}{2}.
\end{aligned}
\label{eq:HC}
\end{equation}
The coefficients of $H_C$ and the parameter $\gamma$ determine the rotation angle of the quantum gates. Due to historical notation custom, a minus sign ahead is added so that the ground state of the Hamiltonian corresponds to the optimal solution of the objective function. In what follows, a state with lower energy (evaluated by $H_C$) is equivalent to a solution with higher score (evaluated by $C$). Design of mixer layer lies in the heart of the algorithm. Here we focus on two types of mixers: $X$-mixers and $XY$-mixers, which will be discussed in next sections.

Taking advantage of superposition state, the cost layer adds different phases to each state according to their energy. In a sense, the cost layers evaluate all solutions simultaneously and store the results as the phases. The mixer layer is the most interesting part where the information is carried from the phase to the amplitude. It amplifies the amplitude according to the phase differences, increasing the probability of low-energy states. In the end of a $p$-level QAOA circuit, measurements are made to extract information from the quantum system. To estimate the superpositioned state $| \phi \rangle$, the quantum computer has to run multiple times to produce a set of measured samples $M=\{(|\vec{x}_i \rangle, f_i) | i = 1, 2, ..., m; f_{i} \leq f_{i+1} \}$, where $\vec{x}_i$ is the bit-wise string of the measurement outcome and $f_i$ is its frequency. Due to probabilistic nature of quantum state, $M$ may include result with significantly small $f_i$. We will drop off those samples whose $f_i < 10\%$, which may also improve the following parameter optimization steps~\cite{barkoutsos2020improving}. A detailed discussion are being given on the drop-off strategy with the experiment results. By classically optimizing the parameters $\vec{\beta}=(\beta _1, ..., \beta _p)$, $\vec{\gamma}=(\gamma _1, ..., \gamma _p)$ with respect to a lost function $F=F(M)$, the probability of low-energy states encoding the optimal solutions and near-optimal solutions will increase. Here the energy expectation is adapted as the lost function: $F = -\frac{1}{m} \sum _{i=1}^{m} p_i C(\vec{x}) $, which is commonly used in various QAOA works due to simplicity. The performance of the algorithm generally depends on the level $p$. We can always increase the frequency of the ground state by adding more levels to QAOA. The probability to observe the ground state is guaranteed to be 100\% when $p \to \infty$~\cite{farhi2014quantum}. Due to limitation of nowadays hardware, the maximum level a quantum computer can support is small. However, QAOA with a finite level has shown noticeable results in various problems~\cite{pagano2020quantum, PhysRevApplied.14.034009, PhysRevX.10.021067, PhysRevResearch.4.033029, PhysRevA.101.012320, cook2020quantum}, including the RNA folding problem under investigated. We set the maximum level to $p_{max} = 8$ for our algorithm and start with $p=2$.

\subsection{$X$-mixers}

In this section we introduce the $X$-mixers QAOA. $X$-mixers are simple mixers mentioned as the prototype~\cite{farhi2014quantum}. Its name comes from the Pauli-$X$ gate used in the mixer layers. The quantum circuits start from a superposition state of all possible solutions $|+\rangle ^{\otimes N}=\sum _{i=1}^{2^N} \frac{1}{\sqrt{2^N}}| i \rangle$, which can be efficiently prepared by $N$ Hadamard gates. Then the cost layers are prepared for a specific RNA sequence. The mixer layer is made up of only $N$ single qubit gates:
\begin{equation}
    U_M(\beta) = e^{i \beta \sum X_i},
\end{equation}
where $X_i$ is Pauli-$X$ gate acting on the $i$-th qubit.

Assuming the stem set $S$ have been calculated classically, the algorithm for X-mixers QAOA goes by the following steps:

1. Map $N$ stems $S=\{ s_i | i=1, 2, ..., N \}$ to a set of qubits $\{ q_i | i=1, 2, ..., N \}$ and calculate all coefficients of $H_C$;

2. Construct a level-2 QAOA circuit with the warm-up parameters $(\vec{\beta}^{w}, \vec{\gamma}^{w})$, and set $p=2$;

3. Optimize the parameters of level-$p$ QAOA to obtain $(\vec{\beta}^{p}, \vec{\gamma}^{p})$ and record the resulting samples $M_{p}$;

4. Examine the sampled results $M_{p}$. If one state gives a frequency larger than 90\%, stop and return the state with the lowest energy among the samples as the approximate solution, otherwise continue with step 5;

5. Use linear interpolation method to generate parameters for level-$p+1$ QAOA. Repeat the parameter optimization process and obtain $(\vec{\beta}^{p+1}, \vec{\gamma}^{p+1})$ and $M_{p+1}$, then set $p$ as $p+1$;

6. Repeat step 4 and 5 until $p=p_{max}$. If no state has a frequency larger than 90\% when the maximum level is reached, stop and return the state with the lowest energy among the samples as the approximate solution.

In step 2, a helpful method is adapted to find near-optimal initial parameters by leveraging parameter concentration. In this paper, parameter concentration is referred as an effect that near-optimal values of $(\vec{\beta}, \vec{\gamma})$ share similar distribution among all instances. It seems a surprising statement at first while it has been mentioned in studies based on both numerical simulation and analytical proof~\cite{PhysRevX.10.021067,PhysRevA.104.L010401,cook2020quantum,Farhi_SKmodel}. We randomly select 20 instances and find their near-optimal parameters by searching exhaustively at level $p=2$. Afterwards the averaged values of these parameters serve as warm-up parameters $(\vec{\beta}^{w}, \vec{\gamma}^{w})$ for all instances. By assuming the concentration phenomenon is applicable to any RNA sequences, we assert $(\vec{\beta}^{w}, \vec{\gamma}^{w})$ are near-optimal for a warm-up. 

In step 5, one starts at layer-$p$ with $2p$ parameters and using linear interpolation method to generate $2(p+1)$ parameters used for layer-$p+1$ circuits. The method is first mentioned by Ref. \cite{PhysRevX.10.021067}, which solves Max-Cut problem with QAOA. It is among one of the parameter optimization methods for deep depth QAOA besides FOURIER interpolation~\cite{PhysRevX.10.021067}, parameters fixing strategy~\cite{lee2021parameters}, bilinear strategy~\cite{lee2022depth}, layerwise training~\cite{PhysRevA.104.L030401}, machine learning~\cite{moussa2022unsupervised}, and so on. Since the parameter concentration is centered in optimization steps, linear interpolation method is more efficient. In this paper, we adapt barycentric interpolation and choose Chebyshev zero modes as the x coordinates. The interpolation go separately to $\vec{\beta}$ and $\vec{\gamma}$. We assign the zero modes $cos\frac{i \pi}{p}$ to each $\beta _i$ ($\gamma _i$) for $i=1,2,...,p$ and apply barycentric interpolation method to obtain parameters for level-$p+1$.

\subsection{Parity-partitioned $XY$-mixers}

\begin{figure}
    \centering
    \includegraphics[width=.7\columnwidth]{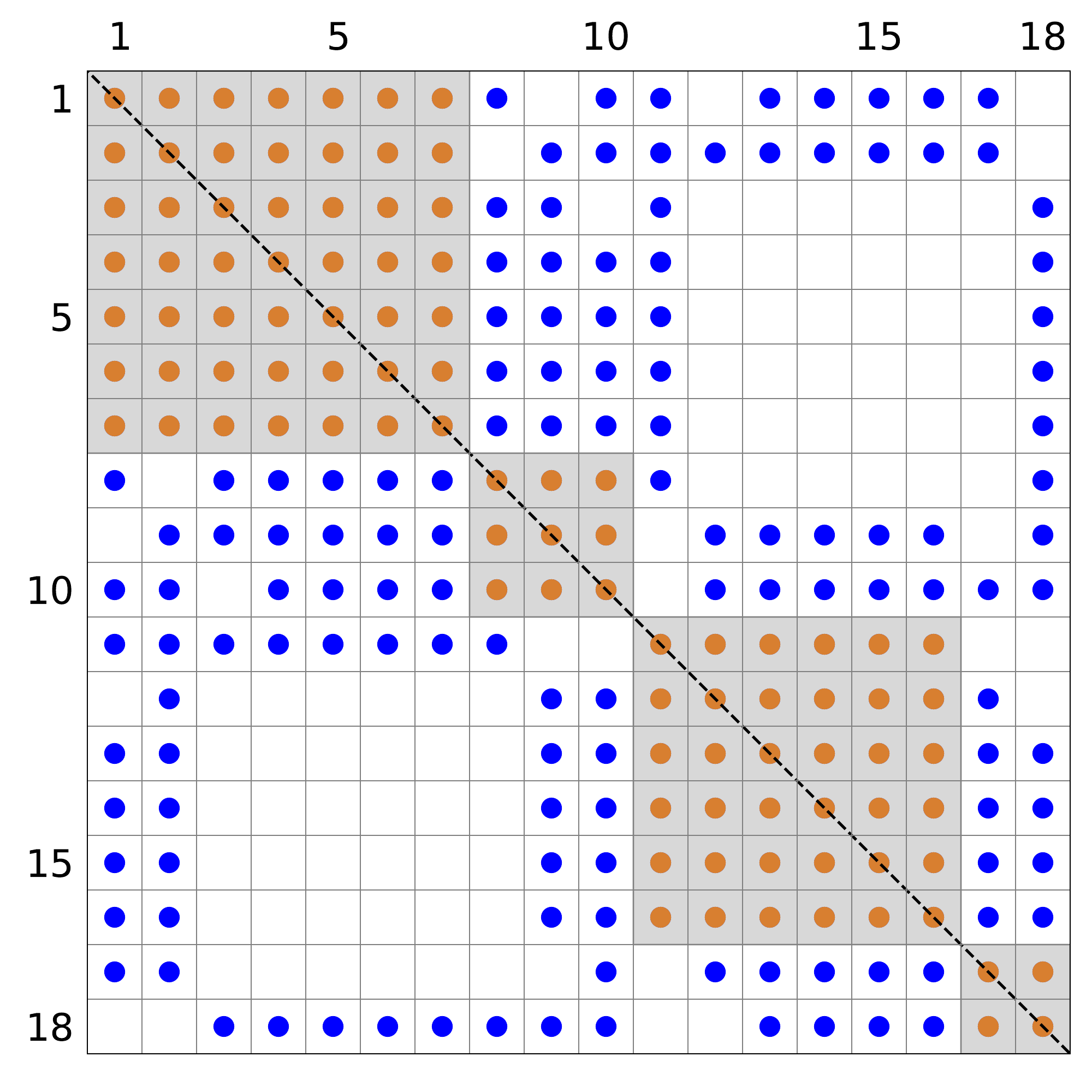}
    \caption{\label{fig:domain} Visualization of the overlap matrix calculated for the sequence PKB092. It contains 18 stems if the minmum stem length is 3. Dot at the $i$-th row and the $j$-th column means the $i$-th stem overlaps the $j$-th stem. Domains are denoted by shaded squares.}
\end{figure}

$X$-mixers have distinguished advantages in circuit design, as their mixer layers require only single qubit gates. However, $X$-mixers is unable to leverage problem characteristics and always tries to search the optimal solution from all $2^N$ solutions. As will be shown in this section, this approach is quite inefficient for RNA folding problem. We hereby introduce the parity-partitioned $XY$-mixers (P-$XY$s). Although the circuit design of P-$XY$s is complex, our simulations suggest that P-$XY$s can potentially produce better results at the same $p_{max}$.

Since the number of stems generally overwhelms the number of bases~\cite{lewis2021qfold}, a pair of arbitrary stems have a high chance to overlap. The fact implies a hidden structure of the QUBO model and helps us to implement P-$XY$s. Consider a set of stems $S=\{ s_i | i=1, 2, ..., N \}$. A matrix $K$ can be used to describe if $s_i$ and $s_j$ are overlapping. In fact, the elements of matrix $K$ are given by Eq. (\ref{eq:kij}). In Fig. \ref{fig:domain} we visualize $K$ of the RNA sequence PKB092 (AAAGUCGCUGAAGACUUAAAAUUCAGG) collected from PseudoBase++ by denoting non-zero elements as colored dots. The shaded squares which are filled with dots along the diagonal are defined as domains. Two arbitrary stems must be overlapping if they belong to the same domain as all the corresponding elements are non-zero. In other words, no more than one stem from one domain must be choose for a non-overlapping folding structure. By introducing an extra zero-length dummy stem for each domain, which stands for ignoring all stems in the corresponding domain, we are restricted to choose exactly one stem from each domain.

It is such single-choice constraint that brings the $XY$-mixers. $XY$-mixers are complex modules but they are spin-invariant if coupled with W-state. Take a 2-qubit system as an example. In this case, W-state is written as $| W \rangle = (|01\rangle + |10\rangle) / \sqrt{2} $. It is a a superposition state of all states where the Hamming weight (number of '1's in the bit-wise string) is exactly one. As the $XY$-mixers act invariant in the subspace spanned by $\{ |01\rangle , |10\rangle \}$, by combining W-state with $XY$-mixers we make sure that the Hamming weight is invariant when performing calculations. For each domain with $d$ stems and one dummy stem, by preparing W-state as initial state and using $XY$-mixers as mixer layers, we are forced to choose exactly one stem. Therefore, the constraint of choosing no overlapping stems is fulfilled. There exist several types of $XY$-mixers~\cite{PhysRevA.101.012320}. P-$XY$s are used in this work because of the simplest circuit structure, which would gain substantial advantages on future experiments.

The algorithm of P-$XY$s QAOA generally follows the steps as $X$-mixers while three main differences are shown below. First of all, the initial state is the superposition state of W state of each domain. Preparing W-state only cost $O(d)$ CNOT gates for a domain with $d$ stems even on a quantum computer of one-dimensional architecture~\cite{PhysRevA.101.012320, PhysRevA.75.032311, PhysRevA.79.042335}. Secondly, the Hamiltonian $H_C$ used in step 1 can be improved. If P-$XY$s are implemented, $Z_i Z_j$ vanishes if $s_i$ and $s_j$ belong to the same domain. Since these quadratic terms are used to penalize overlapping stems, there is no need to introduce extra penalty if the corresponding solutions are expelled from the search space. At last, for a general $XY$-mixers, the mixer layers are:
\begin{equation}
    U_M(\beta) = e^{i \beta \sum_{(i,j)} X_i X_j + Y_i Y_j},
\end{equation}
where the sums are applied to multiple qubit pairs $(i,j)$~\cite{hadfield2019quantum}. In the case of P-$XY$s, the sums range all neighbouring pairs of qubits~\cite{PhysRevA.101.012320}. At last, the warm-up parameters $(\vec{\beta}^{w}, \vec{\gamma}^{w})$ are found different from $X$-mixers while the method to generate them remains the same.

\subsection{Dataset and approximation ratio}

Benchmark RNA sequences in this paper are collected from two datasets: (a) RNA STRAND v2.0~\cite{andronescu2008rna}; (b) PseudoBase++~\cite{taufer2009pseudobase}. The former is a RNA secondary structure and statistical analysis database, containing enormous RNA secondary structures. We focus on the instances provided by PDB database in RNA STRAND, which contains high-resolution ($<$3.5{\AA}) RNA X-ray structures. Besides regular structures, we also particularly collect instances with pseudoknots from PseudoBase++. It is a widely used database of pseudoknot structures, contains over 250 records of pseudoknots obtained in the past 25 years through crystallography, NMR, mutational experiments and sequence comparisons. We select instances with no special bias except for problem size to ensure that the algorithm is evaluated properly. Since simulating a quantum computer using classical computers requires exponential resources, only those that can be captured within 12 qubits by QAOA are studied. To cover more instance, we only keep those stems as long as possible. Since our goal is to maximize the total stem length, in most cases such simplification would not affect the optimal solution of our QUBO model. Note that dropping short stems is not necessary in near-future when quantum hardware would provide sufficient qubits.

Approximate ratio is usually used to evaluate the performance of heuristic algorithms on combinatorial optimization problems. Here we use sensitivity (also referred to as recall in literature) and specificity as approximate ratio~\cite{saito2015precision}. They are both real numbers between 0 and 1 while 1 indicates a perfect prediction. Generally speaking, the sensitivity estimates how many base-pairs the algorithm agree with the reference structure while specificity estimates how many extra base-pairs are presented. Sensitivity is defined as $\frac{TP}{TP + FP}$ where $TP$ is the abbreviation for true positive and $FP$ is for false positive. The positive bases refer to those appearing in the base-pairs given by the algorithm. $TP$ thereby are the number of bases correctly identified by the algorithm, while those which are wrongly predicted as positives are collected by $FP$. Specificity is defined as $\frac{TN}{TN + FN}$ where $TN$ is for true negative and $FN$ is for false negative. In contrary to their counter-part, $TN$ and $FN$ focus on negative bases, namely unpaired bases in the folding structure.

\section{RESULTS}

\subsection{$X$-mixers}

\begin{figure}
    \centering
    \includegraphics[width=\columnwidth]{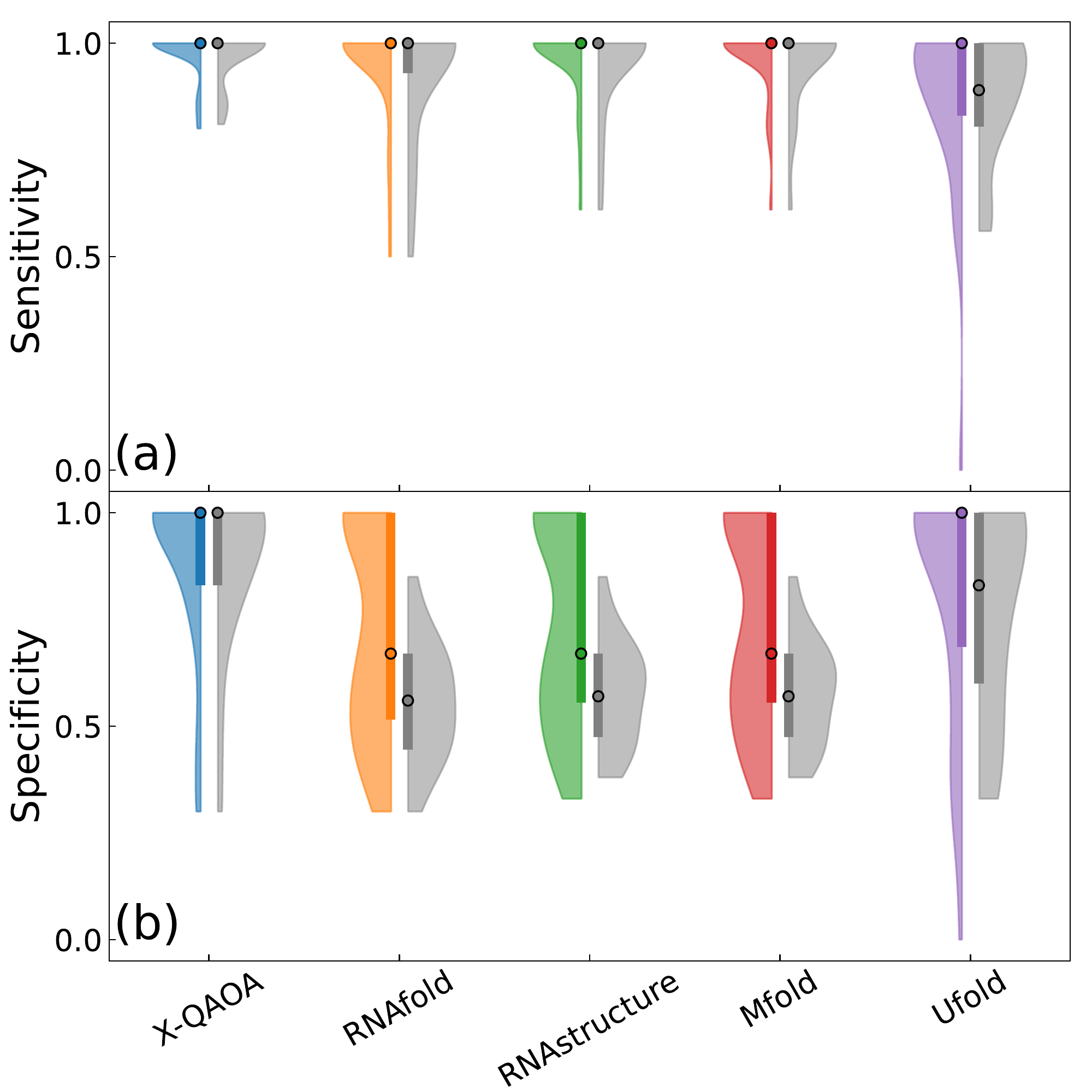}
    \caption{\label{fig:xsum} Violin plots of (a) sensitivity and (b) specificity between QAOA and other classical algorithms tested on the full dataset. The black dots denote medians while the deep colored lines range from first quartile to third quartile of the samples. Grey body on the right hand side represents results of instances with pseudoknots.}
\end{figure}

\begin{table}
\renewcommand{\arraystretch}{1.5}
\centering
\begin{tabular*}{\columnwidth}{lcccc}
\cline{1-5}
Method & Sensitivity & Specificity & Sen. (PK) & Spe. (PK) \\
\cline{1-5}
X-QAOA       & 1.00, 1.00 & 0.83, 1.00 & 1.00, 1.00 & 0.83, 1.00 \\
RNAfold      & 1.00, 1.00 & 0.52, 0.67 & 0.93, 1.00 & 0.45, 0.56 \\
RNAstructure & 1.00, 1.00 & 0.56, 0.67 & 1.00, 1.00 & 0.48, 0.57 \\
Mfold        & 1.00, 1.00 & 0.56, 0.67 & 1.00, 1.00 & 0.48, 0.57 \\
Ufold        & 0.83, 1.00 & 0.69, 1.00 & 0.81, 0.89 & 0.60, 0.83 \\
\cline{1-5}
\end{tabular*}
\caption{\label{table:simu_sum} The first quartile (the value on the left of the comma) and median (the value on the right of the comma) of resulting sensitivity and specificity of all methods tested. Results refined from instances with pseudoknots (PK) are presented on the last two columns.}
\end{table}

In this section, detailed results of $X$-mixers are presented. We use pyqpanda~\cite{pyqpanda} to simulate a noiseless quantum computer where all qubits are connected. Parameter optimization is carrier out by Sequential Least Squares Programming (SLSQP). The performance of QAOA are compared against both energy-based methods (RNAfold, RNAstructure, Mfold) and a newly developed deep learning method (Ufold). The three energy-based methods have become a widely used tools for years thus are good examples for comparison. Ufold is a brand new machine learning architecture and has claimed substantial improvement against various classical algorithms~\cite{fu2022ufold}.

First of all, we investigate with $X$-mixers where the maximum level is 8 and compare the results against classical algorithms. Sometimes QAOA may return several states where they are degenerate, \textit{i.e.}, share the same energy. In these cases, the approximate ratio is given by the mean value of all degenerate states and the ratio may decrease if the extra degenerate states correspond to sub-optimal folding structures. Colored plots (left-half of the violin plots) in Fig. \ref{fig:xsum}(a) and (b) summarize the overall sensitivity and specificity of all methods tested on the whole dataset. The violin plot of sensitivity indicates that our quantum algorithm is comparable to classical methods including machine learning. $X$-mixers, RNAfold, RNAstructure, and Mfold successfully predict the experimental structure for over 75\% instances. For more than a half instances, the results given by QAOA fully captures the base-pairs given by experiments with no extra base-pairs. Detailed values of the first quartile and median is given in Table \ref{table:simu_sum}.

\begin{figure}
    \includegraphics[width=\columnwidth]{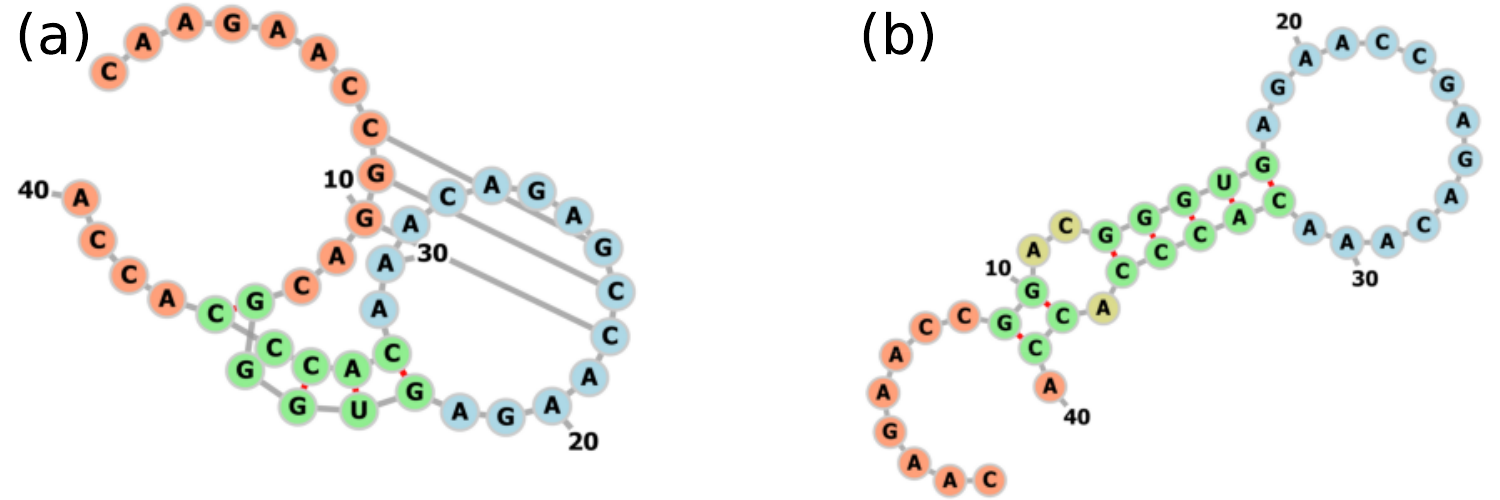}
    \caption{\label{fig:pk_stem} (a) Schematic view of the experiment secondary structure of sequence PKB066 containing pseudoknots. Our algorithm successfully predict the same structure. (b) Secondary structure predicted by the other methods (RNAfold, RNAstructure, Mfold, and Ufold). The visualization is generated by the online tool \textit{forna}~\cite{gendron2001quantitative}.}
\end{figure}

The grey plots in Fig. \ref{fig:xsum}(a) and (b) are results on all instances with pseudoknots. They exhibit another interesting result that our algorithm is surprisingly good at predicting structure of non-nested sequences in comparison to energy-based methods. Because in these sequences containing pseudoknots, the structure with a lower energy is not necessarily the optimal one. Take sequence PKB066 as an example (see Fig. \ref{fig:pk_stem}). Experiment results suggest that the front end sequence would make a pair with the multiloop (bases 8-10 paired with bases 24-22), which actually brings thermodynamic instability and generates misleading results. Therefore in these sequences containing pseudoknots, energy-based dynamic programming methods present notable drawbacks. Comparing with machine learning method Ufold enlightens us with another valuable perspective. The sequence PDB\_01016 has no pseudoknot while Ufold fails to produce any folding structure and outputs a result without any base-pair. Since Ufold is a learning-based method, its performance is inevitably attached to the quality of training data. Such phenomenon is claimed to be related to the incompleteness of its training dataset~\cite{fu2022ufold}. Regarding to these unusual sequences, we can alternatively predict a proper folding structure with QAOA.

\subsection{Experiment results}

\begin{figure*}
    \includegraphics[width=.95\textwidth]{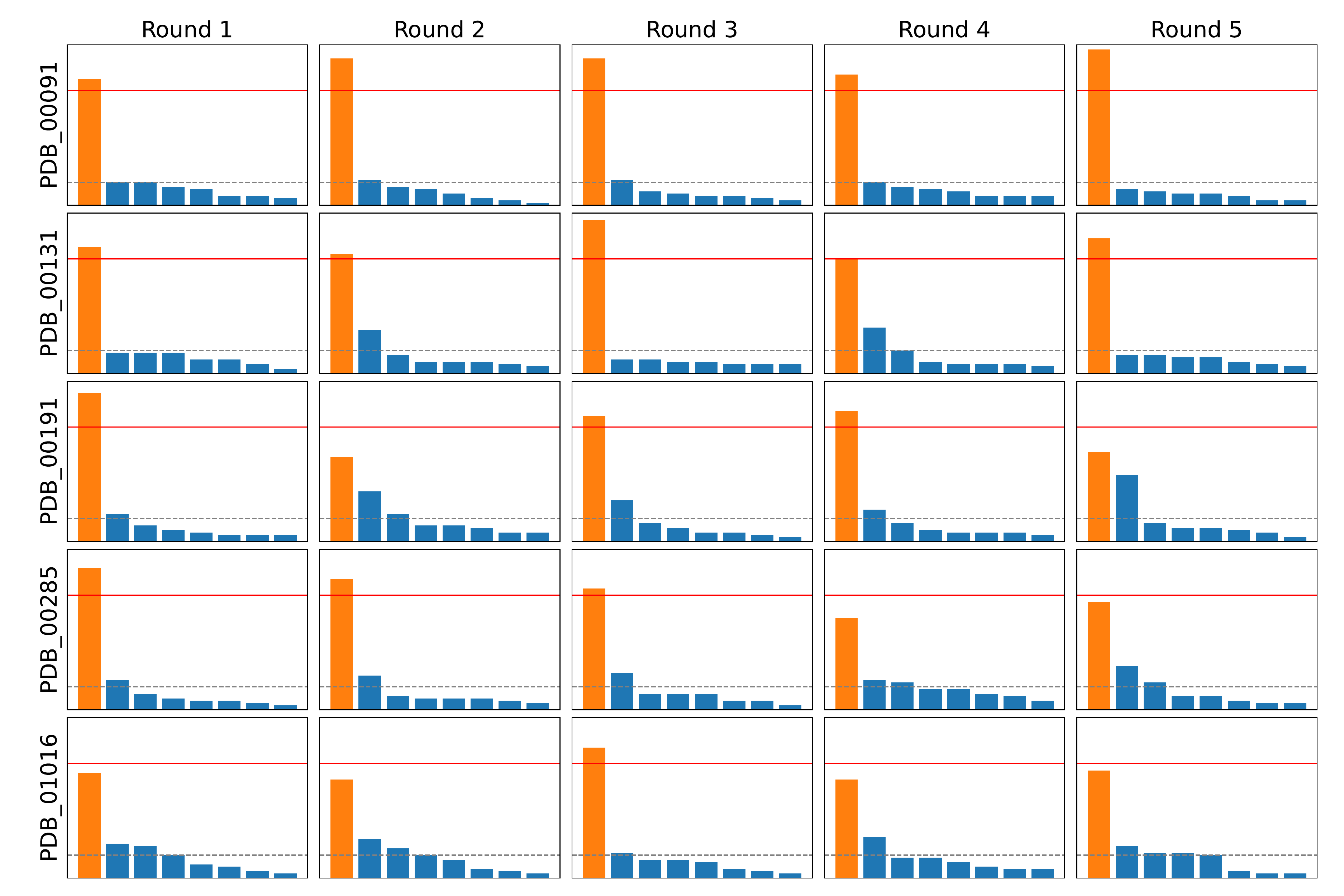}
    \caption{\label{fig:qc_freq} Bar chart of state frequency for all rounds of measurement. Each row corresponds to different instance and each column stands for different round. Inside each subplot, solutions are sorted by the order of their frequency while the bars in orange are the ground states.. The range of frequency is between 0 and 70\% for all subplots. The dashed lines indicate the drop-off value, 10\% and the solid lines denote 50\%.}
\end{figure*}

\begin{table}[b]
\renewcommand{\arraystretch}{1.5}
\centering
\begin{tabular*}{\columnwidth}{lcccccc}
\cline{1-7}
PDB & 1 & 2 & 3 & 4 & 5 & Average \\
\cline{1-7}
00091 & 55\% & 64\% & 64\% & 58\% & 68\% & 62\% \\
00131 & 55\% & 52\% & 67\% & 50\% & 59\% & 57\% \\
00191 & 65\% & 37\% & 55\% & 57\% & 39\% & 51\% \\
00285 & 62\% & 57\% & 53\% & 40\% & 47\% & 52\% \\
01016 & 46\% & 43\% & 57\% & 43\% & 47\% & 47\% \\
\cline{1-7}
\end{tabular*}
\caption{\label{table:qc_sum} The frequency of optimal folding structure (ground state) among the measured samples of five rounds. Frequency of each round is displayed from the 2nd to the 6th column and the average percentage is given in the last column.}
\end{table}

Afterwards, we select instances among the benchmark dataset and run our algorithm on the quantum computer through cloud platform~\cite{qc_clouds}. Running on real quantum chips provides a comprehensive understanding of the algorithm. More importantly, it is a solid and dependable example for numerical results.

Due to hardware limitation, small-size instances which require no more than 4 qubits are investigated. They are: PDB00091, PDB00131, PDB00191, PDB00285, and PDB01016. According to the instruction on the cloud platform, the chip supports 32 CNOT gates at most, which only allows a level-2 QAOA. However, we also find it sufficient for $X$-mixers to produce reliable results on these instances. For all the five RNA sequences, the quantum circuit of our algorithm requires 18 CNOT gates. The number of measurement to obtain one sample set counts for 1000 shots by the default settings of the platform. To avoid communication lag and job queue time, the circuit parameters are fixed as the values in simulation.

In Fig. \ref{fig:qc_freq} we visualize the raw data returned by the cloud platform on which the algorithm for 5 rounds independently for each instances. The critical value for drop-off (10\%) is denoted by dashed line. In other words, those samples whose frequency is below the dashed line would be ignored. With a successful QAOA, this approach is reasonable as these states are likely to have a higher energy. On the other hand, since the ground state (colored in orange) generally has a dominant frequency, our postselection strategy only simplifies the steps to process samples and can hardly improves its performance. It is essential for QAOA to prepare a high-quality state so that we are more likely to hit the optimal and near-optimal states. The algorithm would fail if the frequency of ground state is less than 10\%, in which case these samples are dropped by the postselection.

QAOA shows robust performance by giving stable prediction throughout all examples. Table. \ref{table:qc_sum} shows the frequency of the ground state among the 1000 shots before cut-off. We can see from the table that the average frequency of ground state ranges from 47\% to 62\%, and in most cases we have at least 50\% probability to hit the ground state. Furthermore, among all our test, the ground state (orange color) is always the sample of the highest frequency. We assert that QAOA successfully generates the ground state of $H_C$ and brings reliable prediction of RNA secondary structure for the five instances.

\subsection{$XY$-mixers}

Although the $X$-mixers already show promising results on most instances, we noticed decreased probability of hitting the ground state in some hard instances. Our numerical results suggest that P-$XY$s may offer significant enhancement.

\begin{figure}
    \includegraphics[width=.5\columnwidth]{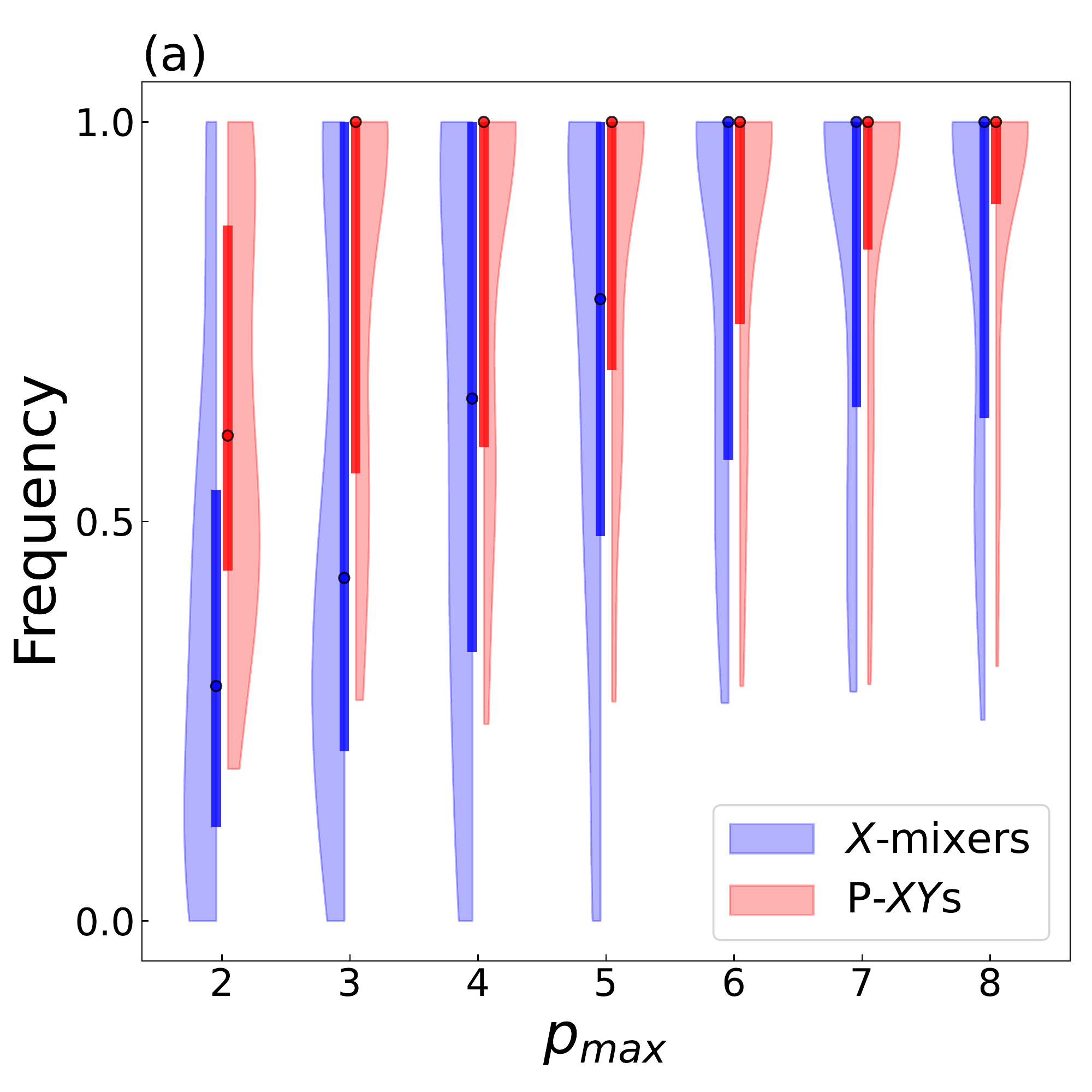}
    \includegraphics[width=.5\columnwidth]{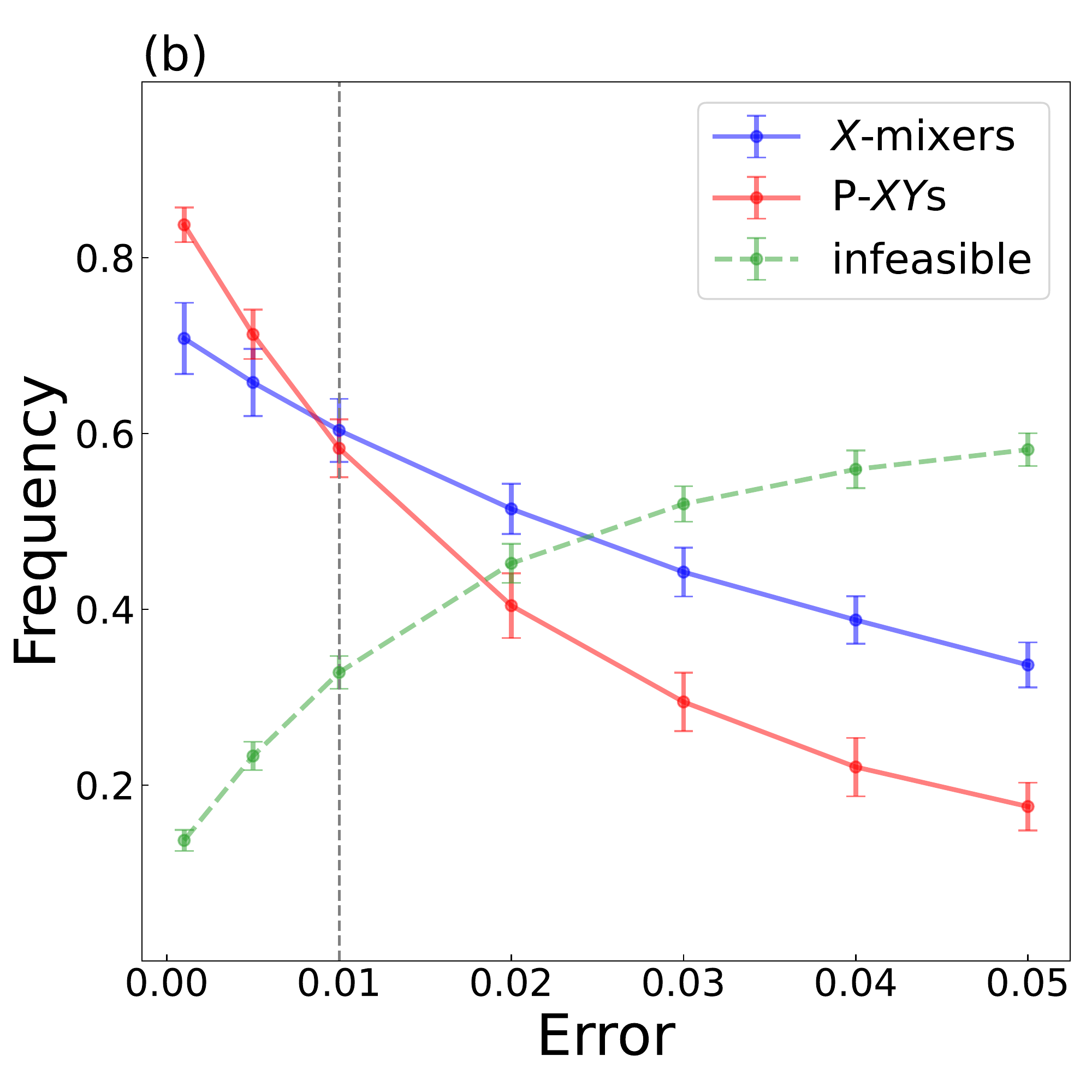}
    \caption{\label{fig:layer} (a) Violin plot of ground state frequency of noiseless simulations. Results of different maximum QAOA level $p_{max}$ is presented along $x$ axis. (b) Frequency of sampling the optimal solution. Results are based on simulation of a noisy quantum computer at different error levels of two-qubit gate. The vertical dashed line suggests the noise level of the experiment device which performs $X$-mixers in Sec. 3.1.}
\end{figure}

We numerically compare the ability to hit the ground state between the two mixers. Fig. \ref{fig:layer}(a) illustrates the results at various $p_{max}$. With the total shots set as 1000, the probability of sampling the ground state increases monotonously with $p_{max}$, which is consistent with performance expectation of QAOA. From the violin plot we see that with level-8 P-$XY$s the chance to measure the ground state is close to 100\% for more than 75\% instances. Averaged frequency of P-$XY$s is about 90.06\% at $p_{max}=8$ while that of $X$-mixers is only 83.32\%. On the other hand, for $X$-mixers the probability to sample the ground state is less than P-$XY$s in all levels presented, especially in the low-level regime. Frequency of hitting ground state with P-$XY$s is 100\% for half of the instances when $p_{max}=3$ meanwhile the $X$-mixers can not guarantee the same performance until $p_{max}=6$.

Next, we simulate a noisy quantum computer to estimate the performance of P-$XY$s on the five instances studied by the experiments in Sec. 3.1. Noise is introduced based on Kraus model and readout assignment errors, which respectively bring errors to quantum gates and grant probability of recording a false measurement outcome. As high-fidelity two-qubit gates remains a bottleneck for practical application of current devices, we focus on performance dependence of errors of two-qubit gates, which only include $CNOT$ gates in our case. The rest parameters of the noise model are chosen empirically to make the simulation results of $X$-mixers closely match the experiment outcomes. As shown in Fig. \ref{fig:layer}(b), we find that P-$XY$s provides noticeable improvement over $X$-mixers at low-level noise regime, where the results have a better frequency of ground state as well as a narrower error bar. However, performance of P-$XY$s decreases more rapidly than $X$-mixers with an increasing gate errors. Our simulation results suggest the performance crossover may happen around an error of 1\%, which is close to current technology levels. As a result, we expect P-$XY$s a promising alternative of $X$-mixers in near future.

We also observe an increasing frequency of infeasible state which violates the Hamming-weight constraints (see dashed line in Fig. \ref{fig:layer}(b)). Such poor performance is a consequence of failing to maintain the problem structure. The improved performance of P-$XY$s greatly depends on the ability of Hamming-weight-preserving operations due to the cooperation between W-state and the mixers, which may greatly reduce the size of space searched by the algorithm. For a sequence with $N$ stems and $N_d$ domains, the $X$-mixers always examine all $2^N$ possible combinations since no constraint is applied on the quantum states. Although P-$XY$s extend the total number of combinations to $2^{N+N_d}$ due to extra qubits, constraints on Hamming weight are applied to each domain, resulting in a total search space of $\prod_{i=1}^{N_d} (d_i+1)$ where $d_i$ is the size of the $i$-th domain. As the condition $d_i+1>2$ is always met, the size of search space in each domain only linearly depends on domain size, which significantly reduces the total search space. In consequence, P-$XY$s are more competent in recognizing ground state and can provide substantial improvement at the same QAOA level.

Last but not least, it is noted that P-$XY$s may be less beneficial for small-size problems. We are going to show the total number of two-qubit quantum gates in the cost layer and mixer layer of P-$XY$s is less than that of $X$-mixers for a large problem size. Consider a domain whose size is $d$. Since any two stems belonging to the same domain are overlapping, there are $(d^2-d)/2$ two-body terms to implement for $X$-mixers. With a general method based on parity of the state~\cite{10.1145/3478519}, it requires $d^2-d$ two-qubit gates in total for the cost layer to implement the non-overlap constraints. On the other hand, although P-$XY$s requires $4(d+1)$ two-qubit gates to implement the mixer layer, the $d^2-d$ gates of the cost layer can be omitted. Thus for $d \geq 6$ the cost of P-$XY$s for each level of QAOA is always less than $X$-mixers. Furthermore, in real devices with a limited qubit connectivity, SWAP gates are needed which would further increase the requirement of $X$-mixers.

\section{DISCUSSION AND CONCLUSION}

In this paper, we propose a QUBO model for RNA folding problem which is solved by a heuristic quantum algorithm, QAOA. Our method is benchmarked on both nested sequences and those with pseudoknots in comparison to four kinds of classical algorithms including energy-based methods (RNAfold, RNAstructure, Mfold) and learning-based method (Ufold). Our simulation results suggest that through QAOA the optimal solution of the problem can be sampled with high probability. It thereby produces with remarkable quality in predicting high-quality folding structure among non-nested sequences compared to traditional energy-based methods. On the other hand, our model archives similar performance to Ufold in most cases. As the training dataset at hand is still limited by experimental techniques, the deep learning method may fail on uncommon instances where QAOA could provide dependable insights. We further implement $X$-mixers QAOA on real quantum computer through cloud computing platform and investigate five sequences using three qubits. It is found that a level-2 QAOA is sufficient to produce optimal solutions for all instances while independent tests suggest the frequency to hit the optimal solution is 47\%  at least and 62\% at most. More importantly, the frequency of ground state is dominant among all measurement outcomes, which accounts for the robust performance of QAOA.

We also show that P-$XY$s can beat the $X$-mixers at the same QAOA level and reach excellent performance for half of all instances at a shallow level $p_{max}=3$ while $X$-mixers requires $p_{max}=6$. Although it remains a challenge to test P-$XY$s on real devices, we try to study the effect of noise by simulating a similar noise environment based on Kraus noise model. Our results suggest P-$XY$s are sensitive to gate errors and their power emerges only in lower noise level, which may be achieved in near-future platforms. The fragility is mainly caused by the breakdown of Hamming-weight constraints under noisy circumstance, where the probability of infeasible states accumulates and irreversibly increase due to gate errors. 

Both simulation and experiment results suggest that quantum algorithms possess great potential in solving practical RNA folding problems. Although the parameter optimization is the most resource-consuming steps, recent works propose an illuminating strategy where the parameters are fixed~\cite{PhysRevA.104.052419}, which suggests that QAOA may bypass the classical optimization steps and thus present evolved efficiency. The parameter transferability is another inspiring strategy~\cite{9605328, shaydulin2022parameter}, which allows the optimal parameters obtained from small size problems to be reused in large size problems. Last but not least, although experimental test of $XY$-mixers is absent in this work, its potential should not stay unnoticed. Both theoretical and experimental studies~\cite{niroula2022constrained, PhysRevA.103.042412} are pushing this ansatz to the edge of realization on various platforms.

In summary, we establish a framework to predict RNA secondary structure using $X$-mixers QAOA and parity-partitioned $XY$-mixers QAOA. By performing numerical simulations, their potential of producing high-quality results on both regular and non-nest sequences are demonstrated. Experiment realization of $X$-mixers is investigated through cloud computing platforms and the algorithm has verified on small scale instances. As the continuous development of quantum hardware, we expect quantum algorithm would provide unparalleled advantages on RNA folding problem.

\section{ACKNOWLEDGEMENTS}
This work was supported by the Innovation Program for Quantum Science and Technology (Grant No. 2021ZD0302300) and the National Natural Science Foundation of China (Grant No. 12034018).

\bibliographystyle{quantum}


\end{document}